\documentclass[aps,10pt,twocolumn,floatfix,longbibliography]{revtex4-1}

\usepackage{amsmath}
\usepackage{amssymb}
\usepackage{bbold}
\usepackage{graphicx}
\usepackage{dcolumn}
\usepackage{bm}
\usepackage{epsfig}
\usepackage[colorlinks=true, citecolor=red, urlcolor=blue, linkcolor=blue]{hyperref}

\usepackage{bookmark}

\begin{document}

\title{Regularized scheme of time evolution tensor network algorithms}

\author{Li-Xiang Cen}
\email{lixiangcen@scu.edu.cn}
\affiliation{Center of Theoretical Physics, College of Physics, Sichuan University, Chengdu 610065, China}
\date{\today}

\begin{abstract}
Regularized factorization is proposed to simulate time evolution for quantum lattice systems.
Transcending the Trotter decomposition, the resulting
compact structure of the propagator indicates a high-order
Baker-Campbell-Hausdorff series.
Regularized scheme of tensor network algorithms
is then developed to determine the ground state energy
for spin lattice systems with Heisenberg
or Kitaev-type interactions. Benchmark
calculations reveal two distinct merits
of the regularized algorithm: it has stable convergence, immune to the bias
even in applying the simple update method to the Kitaev spin liquid;
contraction of the produced tensor network can converge rapidly
with much lower computing cost,
relaxing the bottleneck
to calculate the physical expectation value.
\end{abstract}

\maketitle

Tensor network (TN) and the built numerical algorithms
on it have earned great success in the simulation of quantum
many-body systems \cite{orus2019,cirac2021}, providing deep insights into relevant physics,
e.g., the ground state (GS) property
and dynamics for gapped systems \cite{vidal2004,hast2007,chen2011,scho2011},
the scaling behavior for critical systems \cite{vidal2007ER,even2015,yang2017,bal2017},
and the strongly correlated physics for frustrated
systems \cite{murg2009,yan2011,mezz2012,liao2016,picot2016}.
For one-dimensional (1D) quantum lattice systems, the matrix product state (MPS),
which originally recognized as
the target state of the density matrix renormalization group algorithm \cite{white1992,white1993,ost1995},
is the basis of the infinite time evolving block decimation (iTEBD) algorithm \cite{vidal2007,orus2008}.
The latter allows for direct simulations of both static and dynamic properties of
spin chains in the thermodynamical limit and has boosted vastly the development
of the TN-based algorithm from the $1$D to high-dimensional
lattices
and from the spin system to interacting fermionic
systems \cite{kraus2010,corb2010,poil2014,bult2017}.
The two-dimensional ($2$D) TN state, known as
the projected entangled pair state (PEPS) \cite{vers2004,jor2008} and its variant
the projected entangled
simplex state (PESS) \cite{xie2014}, are natural generalizations of the MPS and
have become standard tools for capturing
the physics of the $2$D strongly correlated quantum systems \cite{jia2008,czar2012,van2016,czar2019}.

Frustrated spin systems can give rise to rich phases of quantum matter
and have attracted intensive interests
in past decades \cite{ander1987,read1991,mars1991,moe2001,wen2002,nor2016}.
Due to strong quantum fluctuations, a particular puzzle
associated with the study of frustrated models, which arouses further the fascination
of wider numerical investigation to them,
is that the results obtained by different methods sometimes cannot reach
consensus---paradigms including the cross-coupled antiferromagnetic
spin ladder \cite{star2004,hung2006,hiki2010,bar2012} and the notorious
kagome antiferromagnet \cite{mars1991,niko2003,singh2007,jia2008PRB,even2010,depen2012,iqbal2013,liao2017,mei2017}.
At this point, the Kitaev honeycomb
model (KHM) \cite{kitaev2006} offers an intriguing
litmus test and research object as well for the numerical methods
in view that: (1) the pure KHM is analytically solvable;
(2) it exhibits gapless and gapped Kitaev
spin liquid phases with fractionalized excitations; (3)
the numerical study is necessarily required when competing interactions,
say, the isotropic Heisenberg interactions and/or the symmetric anisotropic $\Gamma$
interactions are superimposed on the KHM \cite{chal2010,kim2011,ire2014,goh2018,zhang2021}.

Although the variational approach \cite{lee2019} shows that the $2$D TN wave function
can capture precisely the features of the Kitaev spin liquid phase,
previous studies based on the imaginary time evolution TN algorithm
couldn't achieve satisfactory results
for the KHM \cite{ire2014}: the full update infinite PEPS calculation
suffers from a bias with which the GS symmetry
of null magnetization is not guaranteed, while
the simple update method is not able to yield stable convergence from randomly given initial states.
In this paper we will propose a regularized scheme to implement the TN algorithm
in which the time evolution operator is split into a more compact structure
instead of the Trotter-Suzuki formula \cite{suzuki1990}. It turns out that
the regularized TN algorithm is not only able to yield more precise results for
$1$D lattice systems,
but also its $2$D extension can produce reliable non-magnetized outcomes of
spin liquid phases for the KHM.

One computational bottleneck of the $2$D TN algorithm is
the contraction of the full TN which is required in order to obtain
the physical expectation value. Since the approximation of
the contracting process does not meet the variational principle,
the truncation error induced at this step should be much less
than that of the TN wave function so as to warrant the obtained
GS energy to be the upper bound of the exact one.
As this error is visible from the convergence character,
early studies on the PEPS and PESS display that to reach high accuracy
for the contraction is computationally
very expensive \cite{xie2017}.
Remarkably,
for the states produced by the regularized TN algorithm on
the honeycomb lattice with either the Kitaev-type or the Heisenberg
interactions, the accuracy of the contraction is shown to be
dramatically improved even using less computational resources,
which significantly helps to retain the variational principle and
relaxes the bottleneck of the TN algorithm in its application.

Let us start by considering the time evolution operator $e^{-itH}$, or the
Gibbs operator $e^{-\beta H}$, of an infinite
quantum spin chain,
where $H$ is a Hamiltonian with local
interactions and $\beta $ accounts for the inverse temperature. We divide $H$ into $\mathcal{N}$
copies of linked size-$L$ blocks, $H=\sum_{k=1}^{\mathcal{N}}(H_L^{[k]}+v^{k,k+1})$, in
which $v^{k,k+1}$ denotes the interblock coupling and the periodic
boundary condition has been assumed.
The propagator generated by the free terms, i.e., $H_0\equiv \sum_kH_L^{[k]}$
of all disconnected blocks, constitutes a piece-wise local operator $
[U_L(\tau )]^{\otimes \mathcal{N}}$ with $U_L(\tau )=e^{-\tau H_L}$. In the case that
the block size $L$ is considerably large or $\tau $ is small enough, the
whole evolution generated by $H$ can be simulated by the
following decomposition
\begin{equation}
e^{-\tau H}=[\bar{U}_L(\tau )]^{\otimes \mathcal{N}}\times [U_L(\tau )]^{\otimes \mathcal{N}}.
\label{evolu}
\end{equation}
The operator in the second layer, $\bar{U}_L(\tau )$, which
acts on the two coarse-grained sublattices of
$(s_r,s_l)$---the right half of the $k$th block
and the left half of the $(k+1)$th block, is the key ingredient
that the present scheme would outperform the
Trotter decomposition.
It is constructed through an inward algorithm, i.e.,
resorting to the propagators generated by
a pair of length-$L$ blocks that have the same boundary condition, but one has and the other
hasn't the intermediate coupling [see Fig. 1(a)].
Specifically, the two block Hamiltonians with open boundary read as
$H_L=h_l+v^{l,r}+h_r$ and $h_l+h_r$, which give rise to an
``open prescription" of the regularized scheme to construct the operator
\cite{[{Similar prescriptions have been proposed in the regularized numerical
renormalization group in which the pair of block Hamiltonians are
exploited to construct basis states for the compound lattice, see }]rNRG}
\begin{equation}
\bar{U}_L^{(1)}(\tau )=e^{-\tau H_L}e^{\tau (h_l+h_r)}.  \label{coupl1}
\end{equation}
Alternatively, one can make use of the ``periodic prescription", exploiting
propagators generated by two Hamiltonians with connected boundary: $H_L^p\equiv
H_L+v^{r,l}$ and $H_L^{\circleddash}\equiv H_L^p-v^{l,r}$, to construct
\begin{equation}
\bar{U}_L^{(2)}(\tau )=e^{-\tau H_L^p}e^{\tau H_L^{\circleddash}}.
\label{coupl2}
\end{equation}

As the superiority of these two regularized prescriptions will be demonstrated later on
by numerical calculations, the rationality of them can be briefly interpreted by rewriting the
propagator $e^{-\tau H}$ as
\begin{eqnarray}
e^{-\tau H} &=&e^{-\tau H}e^{\tau H_0}e^{-\tau H_0}  \nonumber \\
&=&e^{-\sum_{k=1}^\mathcal{N}V^{k,k+1}}\times e^{-\tau H_0},  \label{rewr}
\end{eqnarray}
in which $V^{k,k+1}$ denotes the collection of terms of
the Baker-Campbell-Hausdorff (BCH) series. It involves the
coupling term $v^{k,k+1}$, the commutator between $v^{k,k+1}$ and $H_0$: $[H,H_0]=
\sum_k[v^{k,k+1},H_0]$, and the resulting nestings. The range of each
$V^{k,k+1}$ will successively increase but be restricted by the order of
the BCH expansion.
If this range does not exceed the length $L$ under a limited expanding order,
the operator of the second layer manifests a piece-wise structure and every local piece
can be simulated by both of $\bar{U}_L^{(1)}(\tau)$ and $\bar{U}_L^{(2)}(\tau)$
\cite{[{See Appendix for the detailed demonstration of the regularized prescription, its generalization
for the honeycomb lattice and the corresponding three-fold staggered iteration}]SupplementalMaterial}.

It is readily seen that the lowest order of the regularized factorization,
the case of $L=2$, recovers the Trotter-Suzuki formula since
the corresponding sublattices have trivial structure and $H_L=v^{l,r}$.
Its high-order scheme with $L\geqslant4$ then
illuminates an improved way to implement the iterative operations for
the iTEBD and,
with its natural generalizations, for the PEPS on high-dimensional
quantum lattice systems.
To be specific,
let $|\psi_0\rangle$ be a randomly given initial state and set the imaginary
time evolution $e^{-\beta H}=(e^{-\tau H})^M$ with
$M\equiv\beta/\tau$ the Trotter number. One applies either the open or
the periodic prescription described in Eqs. (\ref{evolu})-(\ref{coupl2})
to perform $e^{-\tau H}$ iteratively,
so that the projected state $e^{-\beta H}|\psi_0\rangle$ converges to the GS in the limit
$\beta \rightarrow \infty$.
Regularized version of the iTEBD (rTEBD) is readily built in which
the update scheme of the iTEBD is retained
but the representative tensors
$X^{s_l}$ and $Y^{s_r}$ of the MPS [see Fig. 1 (b)] are now defined
on the coarse-grained sites with enlarged spin dimension $d\rightarrow d^{\frac L2}$.

\begin{figure}[t]
\includegraphics[width=0.9\columnwidth]{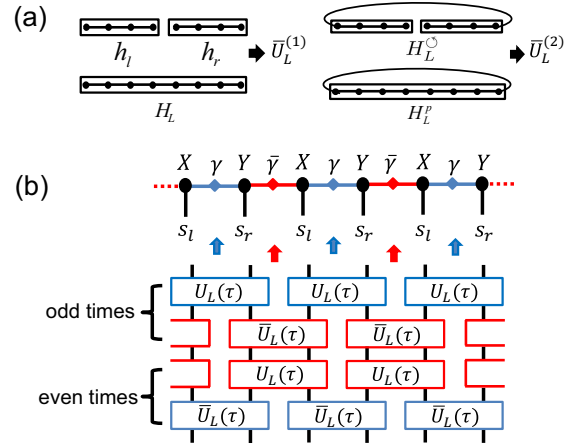}
\caption{Regularized scheme to split the time evolution operator for $1$D
quantum lattices. (a) Block Hamiltonians with different boundary configurations
via which the operators $\bar{U}_L^{(1)}(\tau)$ and
$\bar{U}_L^{(2)}(\tau)$ are constructed [see Eqs. (\ref{coupl1}) and (\ref{coupl2})].
(b) ``Gate" operations of $U_L(\tau)$ and $\bar{U}_L(\tau)$ on
$(s_l,s_r)$ or $(s_r,s_l)$ of the coarse-grained sites, and
a staggered manner to perform the iteration is illustrated.
The local tensor $X$ ($Y$) of the MPS has one physical index
and two virtue bond indices and $\gamma$ ($\bar{\gamma}$)
accounts for the diagonal matrix.}
\end{figure}

In detail, the rTEBD algorithm with various block size $L$ (referred as rTEBD-$L$)
is applied to obtain the GS $|\psi\rangle$
for the infinite $1$D Heisenberg antiferromagnets (HAF)
$H=\sum_i\textbf{S}_i\cdot \textbf{S}_{i+1}$ with $s=\frac 12$
and $s=1$.
%The corresponding GS energy achieved by rTEBD-$L$ with $L=2,4,6,8$
%are shown in Table I.
Below are some particular points worthy to mention.
(1) The algorithm preserves only the translational symmetry of $L$ sites.
A staggered manner to implement the iteration [see Fig. 1 (b)],
that is, imposing the operators
$U_L(\tau)$ and $\bar{U}_L(\tau)$ successively on $(s_l,s_r)$ and $(s_r,s_l)$
in every odd time but imposing the former on $(s_r,s_l)$ and
the latter on $(s_l,s_r)$ in every
even time (recovering the
so-called ``second-order Trotter decomposition" as $L=2$),
is helpful to restore the translational symmetry
of $L/2$-site shifts.
(2) The GS energy per site is given by $e_g=e_L/L$ with
$e_L\equiv\langle \psi|(H_L+v^{k,k+1})|\psi\rangle/\langle \psi|\psi\rangle$.
Numerical calculation displays that the accuracy of $e_L$ is noticeably
better than that of the bulk $\langle H_L\rangle$
and of the coupling $\langle v^{k,k+1}\rangle$, separately.
(3) In all calculations of the rTEBD-$L$ shown in Table I,
the Trotter error is made negligibly small (down to $10^{-8}$).
So it is safe to conclude that the visible improvement of the results obtained
by the high-order rTEBD-$L$ with the same bond dimension $\mathcal{D}$ is not owing
to the reduction of the Trotter error,
but the regularized algorithm takes more
correlations into account hence is able to suppress the truncation error
induced by the approximation of the mean-field-like environment.

\begin{table}[t]
\caption{GS energies of infinite HAF chains given by the rTEBD
algorithm. For references, the exact value of the $s=\frac 12$ model is
$\frac 14-\rm{ln}2\approx-0.4431472$; the value of the $s=1$ model to
the first $12$ digits
is $-1.40148403897$ \cite{white1993}.}
\label{heisen}
\begin{ruledtabular}
\begin{tabular}{ccccc}
~  $e_g$              & rTEBD-2    &   rTEBD-4   & rTEBD-6 &  rTEBD-8 \\ \hline
$       s=1$       &  ~~        &    ~~          &    ~~         &   ~~   \\ \hline
$\mathcal{D}=20$   & -1.4014590 &	-1.4014702	 &	-1.4014740	&	-1.4014764 	 \\
$\mathcal{D}=30$   & -1.4014835	&	-1.4014838	 &	-1.4014838	&	-1.4014839 	 \\ \hline
$    s=\frac 12$   &  ~~        &        ~~      &    ~~         &   ~~   \\ \hline
$ \mathcal{D}=30 $ & -0.4431382 & -0.4431399     &   -0.4431411   & -0.4431418   \\
$ \mathcal{D}=40 $ & -0.4431430 & -0.4431438     &   -0.4431442   & -0.4431446
\end{tabular}
\end{ruledtabular}
\end{table}

The extra cost for the rTEBD with increasing $L$ is the memory resource
that is proportional to the square of the coarse-grained spin dimension $d^{\frac L2}$.
On the other hand, alteration to the time cost should concern comprehensively
the cubic-power relation with the spin dimension $d^{\frac L2}$ and
the relaxation of the step size $\tau$ in the regularized scheme.
Note that the Trotter error $\epsilon$ of a single rTEBD-$L$ iterative step
scales as
$\epsilon\sim c_{L/2}\tau^{\frac L2+1}$ with $c_{L/2}$
the $(L/2)$th-order coefficient of the BCH series. It turns out that the higher
the accuracy required by the outcome, the better the regularized algorithm manifests
its superiority of running speed.
Take the above $s=1$ HAF model as an example. Set the Trotter error $\epsilon \sim 10^{-12}$
which is requested by an output with accuracy of $10\sim 11$ digits [say, the rTEBD-$4$ with $\mathcal{D}=80$
yields $e_g=-1.401484038(1)$].
Time costs of the rTEBD-$4$ and rTEBD-$6$ are reduced to about $1/6$ and $1/2$, respectively,
of that of the rTEBD-$2$, but the rTEBD-$8$ is not able to exhibit speedup until
the Trotter error $\epsilon \lesssim 10^{-14}$.

Extensions of the regularized scheme to diverse configurations of the $2$D
lattice systems are highly nontrivial, among which
we focus below on the honeycomb lattice to elaborate its superiority.
Specifically, we deal with the KHM \cite{kitaev2006} which is defined by
\begin{equation}
H=\frac 12\sum_{\langle i,j\rangle_\gamma}J_\gamma\sigma_i^\gamma \sigma_j^\gamma,
\label{KHM}
\end{equation}
in which the coupling of any two neighboring sites $\langle i,j\rangle$ is dependent
on the direction of their bond $\gamma (=x,y,z)$, as indicated in Fig. 2 (a).
By extending the primitive PEPS \cite{jia2008} to a coarse-grained version,
we apply the regularized factorization of the projection to
obtain the GS at the isotropic points ($J_\gamma=J_{\gamma^\prime}=\pm 1$)
where critical gapless spin liquid phases are formed. As the same scheme is applicable
to the honeycomb lattice with Heisenberg interactions,
the result of the GS of the HAF,
$H=\sum_{\langle i,j\rangle}\vec{S}_i\cdot\vec{S}_j$, will also be presented.

\begin{figure}[t]
\includegraphics[width=0.9\columnwidth]{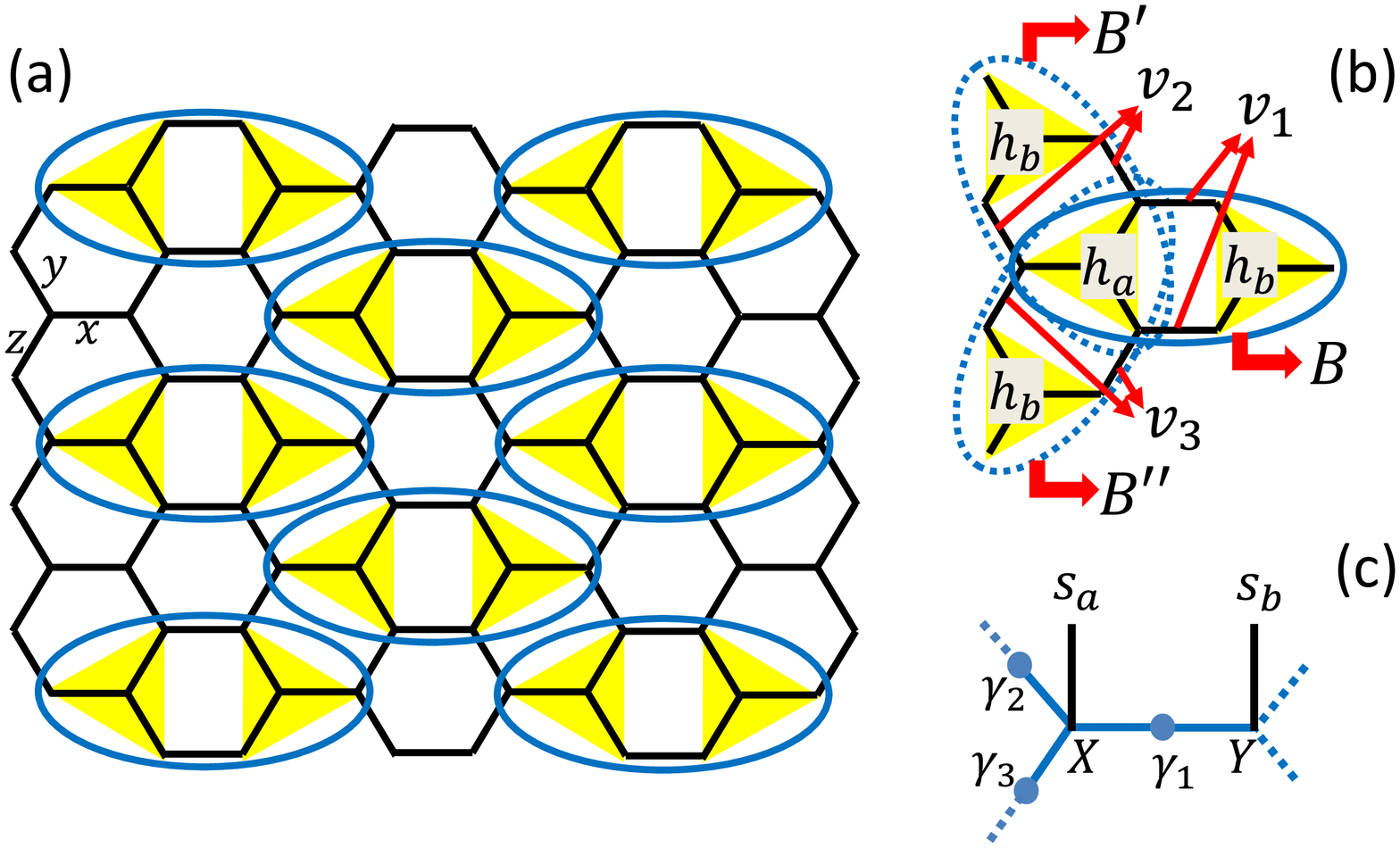}
\caption{Schematic figures for the regularized scheme and
corresponding rTNS-($4,4$) on the $2$D
honeycomb lattice. (a) Partition of the lattice into $8$-site unit
cells $\{B\}$ for both the HAF and KHM. The three $x$, $y$, and $z$ links
are marked out for the KHM [Eq. (\ref{KHM})]. (b)
Unit cells $\{B\}$, $\{B^\prime\}$ and $\{B^{\prime\prime}\}$
along the orientations $x$, $y$, and $z$ on which the operators
$U_1[B]$, $\bar{U}_2[B^\prime]$ and $\bar{U}_3[B^{\prime\prime}]$
are imposed in turn, respectively. (c) Graphical representation of
the rTNS-($4,4$), see Eq. (\ref{2dTNS}).}
\end{figure}

We partition the honeycomb lattice into $\mathcal{N}^2$ copies of $8$-site unit cells $\{B\}$
with $H_B=h_a+v_1+h_b$. The Hilbert space of each $B$ is indicated by the indices $\{s_a,s_b\}$ with
$a$ and $b$ denoting two coarse-grained sublattices [see Fig. 2 (a)]. The total Hamiltonian of the system
hence reads
$H=\sum_{k=1}^{\mathcal{N}^2}(H_B^{[k]}+v_2^{[k]}+v_3^{[k]})$ where $v_2$
and $v_3$ account for the interblock couplings between $B$'s.
The propagator is factorized according to
\begin{equation}
e^{-\tau H}=\bigotimes_{\{B^{\prime\prime}\}}\bar{U}_3[B^{\prime\prime}]\times
\bigotimes_{\{B^\prime\}}\bar{U}_2[B^\prime]\times
\bigotimes_{\{B\}}U_1[B],  \label{evolu2d}
\end{equation}
in which the first layer with $U_1[B]=e^{-\tau H_B}$ denotes the evolution generated by
the free Hamiltonian of all $\mathcal{N}^2$ disconnected $B$'s. The operators
$\bar{U}_2[B^\prime]$ and $\bar{U}_3[B^{\prime\prime}]$ in the second and
the third layers [see Fig. 2 (b)]
are responsible for the interblock couplings $v_2$ and $v_3$, respectively.
They are constructed through the regularized scheme with either the open and
or the periodic prescriptions \cite{SupplementalMaterial}.
The corresponding representation of the wave function, referred as rTNS-($4,4$), is
a coarse-grained version of the PEPS produced by the primitive algorithm \cite{jia2008}:
\begin{eqnarray}
|\Psi\rangle =\sum_{\{s_a,s_b\}}\mathrm{tTr}\prod_{a,b}\gamma _1[B]\gamma _2[B^\prime]\gamma _3[B^{\prime\prime}]X^{s_a}
Y^{s_b}|s_as_b\rangle ,
\label{2dTNS}
\end{eqnarray}
where the product is taken over all $2\mathcal{N}^2$ sublattices $\{a,b\}$ and
$\rm{tTr}$ stands for the tensor trace. The tensors $X^{s_a}$ and $Y^{s_b}$ are of rank four,
possessing the spin indices $s_{a,b}$ with enlarged dimension $2^4$ and three virtual
bond indices of dimension $\mathcal{D}$, and
$\gamma_1 [B]$, $\gamma_2[B']$ and $\gamma_3[B'']$ are diagonal matrices, related to
the singular-value decomposition (SVD) along the three orientations [see Fig. 2 (c)].

\begin{table}[t]
\caption{GS energies of the pure KHM at isotropic points and the HAF obtained
by their rTNS-($4,4$) wave functions with various bond dimensions $\mathcal{D}$ and $\chi$.}
\label{kitaev}
\begin{ruledtabular}
\begin{tabular}{ccccc}
KHM &  ~~~  &  ~~~  &  ~~~  & \\ \hline
$e_g$ & $\chi=\mathcal{D}$ &$\chi=\frac {3\mathcal{D}-1}{2}$ & $\chi=2\mathcal{D}$ & $\chi=3\mathcal{D}$ \\ \hline
$\mathcal{D}=5$   &  -0.3871731 &   -0.3871732         & -0.3871732  &   -0.3871732   \\ \hline
$\mathcal{D}=7$   & -0.3896083	&	-0.3896084	         &	-0.3896084   &-0.3896084 	 \\ \hline
$\mathcal{D}=9$   & -0.3902985	&	-0.3903003	         &	-0.3903004	 & -0.3903005 	 \\ \hline
$\mathcal{D}=11$   & -0.3909176  	&	-0.3909270	         &	-0.3909290	 & -0.3909293 	 \\ \hline \hline
HAF &  ~~~  &  ~~~  &  ~~~  & \\ \hline
$e_g$     & $\chi=\mathcal{D}$       &$\chi=\frac 32\mathcal{D}$ & $\chi=2\mathcal{D}$  &   $\chi=3\mathcal{D}$ \\ \hline
$\mathcal{D}=6$   &  -0.54123282  &  -0.54123284  &  -0.54123284    &  -0.54123284    \\ \hline
$\mathcal{D}=8$   &  -0.54384911  &  -0.54384959  &  -0.54384974    &  -0.54384975    \\ \hline
$\mathcal{D}=10$  &  -0.54412707	 &  -0.54412974	 &	-0.54413015   &  -0.54413018    \\ \hline
$\mathcal{D}=12$  &  -0.54425157	 &  -0.54425354	 &	-0.54425368	  &  -0.54425370
\end{tabular}
\end{ruledtabular}
\end{table}

We first run the regularized algorithm for the iterative imaginary time evolution incorporating
with the simple update method to obtain the GS of the system in the thermodynamic
limit. In view of the symmetry of the lattice with respect to the spatial $120^{\rm{o}}$-rotation, a three-fold
staggered way to perform the operators is exploited for the iteration \cite{SupplementalMaterial}.
We take $\tau =0.1$ initially and reduce it successively till $\tau=0.1\times 2^{-6}$.
The randomly generated initial states converge to a stable rTNS-($4,4$)
almost in all times, evidenced by the (normalized) singular values
of $\gamma_1[B]$, $\gamma_2[B^\prime]$ and $\gamma_3[B^{\prime\prime}]$.
The algorithm outputs the same results for the ferromagnetic and anti-ferromagnetic KHMs ($J_\gamma=\pm1$)
including the values of $\gamma_{1,2,3}$ and the GS energy,
and their accumulated iteration
number should be much larger ($\sim 20$ times or more) than that of the HAF model.
On the other hand, the periodic prescription is shown to perform better
than the open one, and this becomes dramatic for the KHM with increasing $\mathcal{D}$:
the convergence by the open prescription becomes difficult
for the KHM with $\mathcal{D}>6$, but the periodic one can work for much higher $\mathcal{D}$.

As has been mentioned previously, the regularized algorithm can reduce two kinds
of error, the Trotter decomposition and the approximation of the mean-field-like
environment with respect to the simple update. Both these two aspects
should play decisive roles in achieving stable convergence for the KHM.
This is because that the ``evolutionary rate" caused by the imaginary-time propagator
of the gapless system is very slow and the tendency toward the optimal target state
is easily affected by the two kinds of errors. This is also the reason why the primitive
PEPS algorithm fails to obtain satisfactory results:
the two-site propagator together with the mean-field-like environment of
the simple update scheme is too rough to obtain the strongly
correlated spin liquid phase
although the Trotter error could be made negligibly small;
while in the full update scheme the Trotter error is relatively large as the
time cost of the algorithm is very expensive and increases linearly
with $1/\tau$.

With the obtained wave function, we can then estimate physical expectation values
by applying the contraction algorithm to the full summation over the $2$D TN.
It turns out that the contraction associated with the rTNS-($4,4$) has desirable convergency,
e.g., the boundary MPS method \cite{jor2008} is able to yield sufficient accuracy
by taking the bond dimension $\chi$ to be $2\mathcal{D}\sim3\mathcal{D}$.
Moreover, a notable physical property of the rTNS-($4,4$) obtained for the KHM
is its null magnetization for all $\mathcal{D}$ values shown in Table II
($\langle\sigma_i^\gamma\rangle$ being $10^{-12}\sim10^{-10}$ attributed to
the numerical error),
which unambiguously affirms the outputted spin liquid phase.
The average values of the GS energy, i.e.,
$e_g\equiv\frac {1}{8}(\langle H_B\rangle+\langle v_2\rangle+\langle v_3\rangle)$ per site
or $\frac 23e_g$ per bond, are shown in Table II for both the KHM and HAF.
The outputs of the contraction converge rapidly with $\chi$ (slightly slowing down as $\mathcal{D}$ increases).
When $\chi$ takes the value of $2\mathcal{D}\sim3\mathcal{D}$, the accuracy of the contraction
is already above $10^{-6}$ or $10^{-7}$ for the KHM and the HAF, respectively, which
is sufficient to warrant the output to be the upper bound of the exact energy.
The converged value of $\mathcal{D}=13$ for the KHM is $e_g=-0.39134$, which is
about $0.58\%$ higher than the exact one as the gapless GS energy
exhibits algebraic convergence with $\mathcal{D}$.
For the HAF model, the result of $\mathcal{D}=14$ ($e_g=-0.54432$,
attainable by a desktop or laptop) is already comparable to those of the Monte Carlo ($-0.54455$) \cite{low2009}
and of the second renormalization on the primitive TN state with $\{\mathcal{D},\chi\}=\{16,130\}$ ($-0.54440$) \cite{xie2009,zhao2010},
and an accuracy of $10^{-6}$ ($\mathcal{D}\approx 24$ according to extrapolation) is achievable by the current
computing power.

To summarize, the regularized TN algorithm is developed and shown to be able
to yield reliable results for both
$1$D and $2$D quantum lattice systems with modest computational resources.
As the ability to capture
the spin liquid phase is tested by the pure KHM with non-magnetized outputs,
further applications of the algorithm to the KHM with competing interactions
and to other frustrated models
are expected and should be addressed elsewhere. The lattice units we have
adopted for the honeycomb lattice assume a simple structure with two sublattices.
Proper partitions of the honeycomb and other lattices into unit cells with multiple
components will result in regularized TN states with multiple local tensors on which
the high-order SVD is suitably applied in order to implement the update.
Since the factorization method for the time evolution is applicable to
general quantum lattices, systematic analyses and applications of the
regularized algorithm to probe the GS as well as the dynamical and thermodynamic
properties for varieties of lattice systems, will be a research subject of the next step.

\textit{Acknowledgments---} This work was supported by the NSFC, China, under Grant No. 12147207.

\bibliography{rTN}

\clearpage

\appendix

\section{Regularized factorization of the propagator and the limited order of the BCH expansions}

In order to explain the rationality for the factorization of the propagator shown in Eqs. (\ref{evolu})-(\ref{coupl2}),
we first reveal that the combining operator $e^{-\tau H}e^{\tau H_0}$ given in Eq. (\ref{rewr})
possesses an effective piece-wise structure. To this end, one applies
the BCH expansion
\begin{equation}
e^{-\tau H}e^{\tau H_0}=e^{-\tau \tilde{V}_1-\frac {\tau^2}2 \tilde{V}_2+\frac {\tau^3}{12} \tilde{V}_3+\cdots},
\end{equation}
in which $\tilde{V}_i$'s contained in the first three terms of the exponential read as
\begin{eqnarray}
\tilde{V}_1&=&H-H_0=\sum_k v^{k,k+1}\equiv\sum_k\tilde{v}_1^{[k]}, \\
\tilde{V}_2&=&[H,H_0]=\sum_k[v^{k,k+1},H_0]\equiv\sum_k\tilde{v}_2^{[k]}, \\
\tilde{V}_3&=&\sum_k [H+H_0,[v^{k,k+1},H_0]]\equiv\sum_k\tilde{v}_3^{[k]}.
\end{eqnarray}
Truncation to the first term of the above BCH series yields simply the conventional Trotter-Suzuki decomposition.
It is crucial to note that
the commutation and the corresponding nestings involving $v^{k,k+1}$ in high-order terms
will expand the range of the coupling (e.g., for the case that $H$ involves only the
nearest-neighboring interaction, it widens two more sites by increasing each
order of the expansion) before it overlaps the adjacent ones. To guarantee the
effective piece-wise structure of the operator,
the length $L$ of the block and the expanding
order $n$ of the BCH series should satisfy $L/2\geq n$, e.g., for the case of the nearest-neighboring
interaction. This is clearly seen from the fact that the term $V^{k,k+1}$ in Eq. (\ref{rewr})
can be expressed explicitly as
\begin{equation}
V^{k,k+1}=\tau\tilde{v}_1^{[k]}+\frac {\tau^2}2\tilde{v}_2^{[k]}-\frac {\tau^3}{12}\tilde{v}_3^{[k]}+\cdots,
\label{vkk}
\end{equation}
in which all $\tilde{v}_n^{[k]}$'s ($n=1,\cdots,L/2$) are of local form, e.g.,
\begin{eqnarray}
\tilde{v}_1^{[k]}&=&v^{k,k+1}, \\
\tilde{v}_2^{[k]}&=&[\tilde{v}_1^{[k]},h_r^{[k]}+h_l^{[k+1]}], \\
\tilde{v}_3^{[k]}&=&[2h_r^{[k]}+2h_l^{[k+1]}+v^{k,k+1},\tilde{v}_2^{[k]}]].
\end{eqnarray}
As the range of every $V^{k,k+1}$ is limited by the expanding order,
the condition $L/2\geq n$ warrants that $V^{k,k+1}$ is local and
satisfies $[V^{k,k+1},V^{k^\prime,k^\prime+1}]=0$.

To demonstrate the validity of the decomposition one then needs only to show that every piece of the operator, $e^{-V^{k,k+1}}$,
can be efficiently simulated by the two prescriptions of $\bar{U}_L(\tau)$ presented in
Eq. (\ref{coupl1}) and (\ref{coupl2}), alternatively. This can be recognized directly by expanding them via the BCH
formula, both of which give rise to
\begin{equation}
\bar{U}_L(\tau)=e^{-\tau v_{lr}-\frac {\tau ^2}2[v_{lr},h_l+h_r]+\frac
{\tau^3}{12}[2h_l+2h_r+v_{lr},[v_{lr},h_l+h_r]]+\cdots}.
\end{equation}
Under the restriction of the expanding order $n\leq L/2$, the above expression reproduces
exactly the one of $e^{-V^{k,k+1}}$ except for a translation of $L/2$ sites.

In the decomposing scheme of Eq. (\ref{evolu}), we have set the ingredient of the propagator generated by the free
Hamiltonian term to be the first layer. On the contrary, an alternative way to decompose
the propagator can also be given by rewriting Eq. (\ref{rewr}) as
\begin{eqnarray}
e^{-\tau H} &=&e^{-\tau H_0}e^{\tau H_0}e^{-\tau H}  \nonumber \\
&=&e^{-\tau H_0}\times e^{-\sum_{k=1}^\mathcal{N}\bar{V}^{k,k+1}},  \label{rewra}
\end{eqnarray}
in which $\bar{V}^{k,k+1}$, yielded by the BCH expansion of $e^{\tau H_0}e^{-\tau H}$,
possesses a similar local structure with $V^{k,k+1}$ shown in Eq. (\ref{vkk}).
It thus leads to
\begin{equation}
e^{-\tau H}=[U_L(\tau )]^{\otimes \mathcal{N}}\times[\tilde{U}_L(\tau )]^{\otimes \mathcal{N}}.
\label{evolu2}
\end{equation}
The corresponding open and periodic prescriptions to construct $\tilde{U}_L(\tau )$ are
given by
\begin{equation}
\tilde{U}_L^{(1)}(\tau )=e^{\tau (h_l+h_r)}e^{-\tau H_L}  \label{coupla1}
\end{equation}
and
\begin{equation}
\tilde{U}_L^{(2)}(\tau )=e^{\tau H_L^{\circleddash}}e^{-\tau H_L^p},
\label{coupla2}
\end{equation}
respectively. That is to say, the factorization scheme of Eq. (\ref{evolu2}) exchanges the order of the two layers
of the operations indicated in Eq. (\ref{evolu}) and the associated prescriptions (\ref{coupl1}) and (\ref{coupl2})
responsible for the interblock coupling should also change the order of their generating operators
accordingly [cf. expressions of $\bar{U}_L^{(1,2)}(\tau)$ presented in Eqs. (\ref{coupl1}) and (\ref{coupl2})].
At this stage, a different version of the staggered way to implement the action of $e^{-\tau H}$
for the iteration, i.e., via the decomposition of Eq. (\ref{evolu})
at every odd time and via that of Eq. (\ref{evolu2}) at every even time, is suggested.
Since it adopts a distinct strategy from that shown in Fig. 1(b), a compatible iterative scheme can be designed
by combining these two staggered strategies, via which the Trotter error can be further suppressed.

\section{Regularized factorization and staggered iteration
for the honeycomb lattice}

\begin{figure}[t]
\includegraphics[width=0.85\columnwidth]{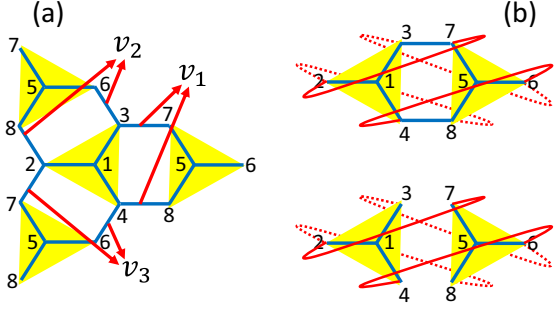}
\caption{Schematic figures of the block Hamiltonians employed
to construct the propagator in the regularized TN algorithm.
(a) An example to mark the order of sites in the lattice units.
(b) The pair of block Hamiltonians with connected boundary,
$H_B^p=H_B+v_2+v_3$ (upper panel) and $H_B^\circleddash=H_B^p-v_1$ (bottom panel),
with which the operator $\bar{U}_1[B]$ is constructed.}
\end{figure}

\begin{figure}[t]
\includegraphics[width=0.85\columnwidth]{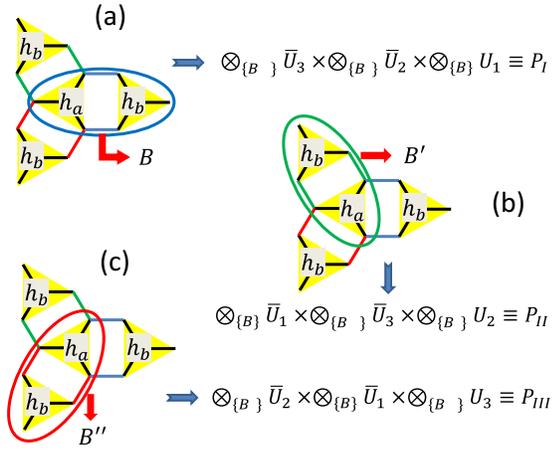}
\caption{Schematic figures of the three different ``gate" sequences to simulate
the propagator generated by the honeycomb lattice system. The notations $P_I$, $P_{II}$
and $P_{III}$ are introduced to specify them of (a), (b) and (c) which relate to each other by a $120^{\rm{o}}$
spatial rotation in turn.}
\end{figure}

Since the Hamiltonians utilized in the open prescription to construct the propagator
are clear themselves, we illustrate here the Hamiltonians with connected boundary
which are employed by the periodic prescription to construct
the propagator. By choosing the lattice units $\{B\}$ for the free
Hamiltonian $H_0=\sum_{k=1}^{\mathcal{N}^2}H_B^{[k]}$, one needs
to construct $\bar{U}_2[B^\prime]$ and $\bar{U}_3[B^{\prime\prime}]$
for the factorization shown in Eq. (\ref{evolu2d}). The periodic prescription
for $\bar{U}_2[B^\prime]$ exploits a block Hamiltonian with periodic
boundary condition
$H_B^p=H_B+v_2+v_3$ (noticing $H_{B^\prime}^p=H_{B^{\prime\prime}}^p=H_{B}^p$, see Fig. 3)
and the other one by subtracting the term $v_2$ from $H_B^p$: $H_{B^\prime}^\circleddash=H_{B^\prime}^p-v_2$,
that is,
\begin{equation}
\bar{U}_2[B^\prime]=e^{-\tau H^p_{B^{\prime}}}e^{\tau H_{B^\prime}^\circleddash}.
\end{equation}
The operator $\bar{U}_3[B^{\prime\prime}]$ is constructed similarly
\begin{equation}
\bar{U}_3[B^{\prime\prime}]=e^{-\tau H^p_{B^{\prime\prime}}}e^{\tau H_{B^{\prime\prime}}^\circleddash},
\end{equation}
in which $H_{B^{\prime\prime}}^\circleddash=H^p_{B^{\prime\prime}}-v_3$.

Concerning the symmetry of the
lattice with respect to the spatial $120^{\rm{o}}$-rotation,
one can also choose the lattice units $\{B^\prime\}$ or $\{B^{\prime\prime}\}$
as the free Hamiltonian, ie., $H_0=\sum_{k=1}^{\mathcal{N}^2}H_{B^\prime}^{[k]}$
or $H_0=\sum_{k=1}^{\mathcal{N}^2}H_{B^{\prime\prime}}^{[k]}$.
Accordingly, the decomposition of the propagator $e^{-\tau H}$
can be realized by the following two different ``gate" sequences
\begin{eqnarray}
e^{-\tau H}&=&\bigotimes_{\{B\}}\bar{U}_1[B]\times
\bigotimes_{\{B^{\prime\prime}\}}\bar{U}_3[B^{\prime\prime}]\times
\bigotimes_{\{B^{\prime}\}}U_2[B^{\prime}], \label{sfact1} \\
e^{-\tau H}&=&\bigotimes_{\{B^{\prime}\}}\bar{U}_2[B^{\prime}]\times
\bigotimes_{\{B\}}\bar{U}_1[B]\times
\bigotimes_{\{B^{\prime\prime}\}}U_3[B^{\prime\prime}].
\label{sfact2}
\end{eqnarray}
Here, $U_2[B^{\prime}]$ and $U_3[B^{\prime\prime}]$ in the first layer
of the two sequences are just the propagators generated by the two free Hamiltonians
defined on $\{B^\prime\}$ and $\{B^{\prime\prime}\}$, respectively.
The operator $\bar{U}_1[B]$ is responsible for the coupling $v_1$
and is constructed via $\bar{U}_1[B]=e^{-\tau H_B^p}e^{\tau H_B^\circleddash}$
with $H_B^\circleddash=H_B^p-v_1$.
These three different gate sequences to simulate $e^{-\tau H}$, i.e., indicated
by Eqs. (\ref{evolu2d}), (\ref{sfact1}) and (\ref{sfact2}), relate to each other by a $120^{\rm o}$ spatial rotation in turn
and are shown schematically in Fig. 4 with the notations $P_{I}$, $P_{II}$ and
$P_{III}$, respectively. In analogy to the staggered iteration previously proposed
for the $1$D lattice system (cf. Fig. 1), a $3$-fold staggered way to apply these gate sequences
can be devised to implement the iteration in the regularized TN algorithm,
which is helpful to reduce further the error of the factorization and restore the $120^{\rm{o}}$
rotational symmetry for the honeycomb lattice.

In the practical performance of the iteration, these three sequences of gate operations
are implemented following
the order of $P_I$, then $P_{III}$, and $P_{II}$ last. The benefit of doing so is that the consecutive twice
operations imposed on the same lattice units, e.g., $\bar{U}_3[B^{\prime\prime}]$ in the last layer
of $P_I$ and $U_3[B^{\prime\prime}]$ in the first layer of $P_{III}$ that are imposed on the same
$\{B^{\prime\prime}\}$, can be merged, which is able to
save $1/3$ of the iteration times.

\end{document}